\documentclass[a4paper,11pt]{article}
\pdfoutput=1 

\usepackage{jcappub, float, caption, subcaption, bm, txfonts, lmodern, multirow} 
\usepackage[sort&compress,numbers]{natbib}
\usepackage{calc, amsmath} 
\usepackage{soul}

\usepackage[T1]{fontenc} 

\newcommand{\s}{\,{\rm s}}
\newcommand{\m}{\,{\rm m}}
\newcommand{\Kel}{\,{\rm K}}
\newcommand{\eV}{\,{\rm eV}}
\newcommand{\MeV}{\,{\rm MeV}}
\newcommand{\GeV}{\,{\rm GeV}}
\newcommand{\Hz}{\,{\rm Hz}}
\newcommand{\nHz}{\,{\rm nHz}}
\def\mg{m_{\rm g}}
\def\kB{k_{\rm B}}
\def\kf{k_{\rm f}}
\def\hc{h_{\rm c}}
\def\dd{{\rm d}}
\def\ii{{\rm i}}
\def\Sp{{\rm Sp}}

\def\BB{\bm B}
\def\JJ{\bm J}
\def\ee{\bm e}
\def\ff{\bm f}
\def\kk{\bm k}
\def\uu{\bm u}
\def\xx{\bm x}
\def\eq{\equiv}
\def\ij{{ij}}
\def\ijl{{ijl}}
\def\lsim{\lesssim}
\def\d{\delta}
\def\sig{\sigma}
\def\eps{\epsilon}
\def\Lam{\Lambda}
\def\gam{\gamma}

\def\munu{{\mu\nu}}
\def\lambar{\lambdabar}

\def\aaa{\mathcal{A}}
\def\RHO{\rho}
\def\eee{\mathcal{E}}
\def\lll{\mathcal{L}}
\def\fff{\mathcal{F}}
\def\nnn{\mathcal{N}}
\def\uuu{\mathcal{U}}
\def\ssf{\mathsf{S}}
\def\bra{\langle}
\def\ket{\rangle}
\def\K{{\rm K}}
\def\M{{\rm M}}
\def\Re{{\rm Re}}
\def\GW{{\rm GW}}
\def\EH{{\rm EH}}
\def\EW{{\rm EW}}
\def\TT{{\rm TT}}
\def\rms{{\rm rms}}

\def\sat{{\rm sat}}
\def\QCD{{\rm QCD}}
\def\mat{{\rm mat}}
\def\rad{{\rm rad}}
\def\cut{{\rm cut}}
\def\cutz{{\rm 0}}
\def\diag{\rm diag}
\def\phys{{\rm phys}}
\def\crit{{\rm crit}}

\newcommand{\Del}{\bm\nabla}
\newcommand{\DDel}{\bm\nabla^2}

\newcommand{\Deldot}{\bm\nabla\cdot}
\newcommand{\Delcrs}{\bm\nabla\times}
\newcommand{\p}{\partial}

\def\pfrac#1#2{\frac{\partial#1}{\partial#2}}

\def\Dfrac#1#2{\frac{D#1}{D#2}}

\newcommand{\Fig}[1]{figure~\ref{#1}}
\newcommand{\Sec}[1]{section~\ref{#1}}

\newcommand{\Tab}[1]{table~\ref{#1}}

\title{Tensor spectrum of turbulence-sourced gravitational waves as a constraint on graviton mass}

\author[a,b]{Yutong He,}
\author[a,b,c]{Axel Brandenburg,}
\author[c]{and Aditya Sinha}

\affiliation[a]{Nordita, KTH Royal Institute of Technology and Stockholm University,\\Hannes Alfv\'ens v\"ag 12, 10691 Stockholm, Sweden}
\affiliation[b]{Department of Astronomy, AlbaNova University Center, Stockholm University\\10691 Stockholm, Sweden}
\affiliation[c]{Department of Physics, Carnegie Mellon University, Pittsburgh, PA 15213, USA}

\emailAdd{yutong.he@su.se}
\emailAdd{brandenb@nordita.org}
\emailAdd{asinha5@andrew.cmu.edu}

\abstract{
We consider a generic dispersive massive gravity theory and numerically
study its resulting modified energy and strain spectra of tensor gravitational waves (GWs)
sourced by (i) fully developed turbulence during the electroweak phase
transition (EWPT) and (ii) forced hydromagnetic turbulence during
the QCD phase transition (QCDPT).
The GW spectra are then computed in both spatial and temporal Fourier domains.
We find, from the spatial spectra, that the slope modifications are
weakly dependent on the eddy size at QCDPT, and, from the temporal
spectra, that the modifications are pronounced in the $1$--$10\nHz$ range
-- the sensitivity range of the North American Nanohertz Observatory
for Gravitational Waves (NANOGrav) -- for a graviton mass $\mg$ in the
range $2\times10^{-23}\eV\lsim \mg c^2\lsim7\times10^{-22}\eV$.
}


\arxivnumber{}

\begin{document}
\maketitle
\flushbottom

\section{Introduction}
\label{sec:intro}

The history of gravity theories alternative to general relativity
(GR) is almost as long as that of GR itself.
GR propagates massless gravitons described by a linear dispersion
relation $\omega = ck$, where $c$ is the speed of light.
One possibility of modifying GR is by having a nonlinear dispersion
relation of the form $\omega = \sqrt{c^2k^2 + \omega_\cutz^2}$, where
$\omega_\cutz^2 = (\mg c^2/\hbar)^2 + \aaa c^\alpha k^\alpha$ is a
frequently adopted form of a nonlinear modification \cite{Mirshekari+11,
Cembranos+18}.
Here $\mg c^2/\hbar$ is an effective mass term for a nonzero
graviton mass $\mg$ \cite{Lee14}, and $\mathcal{A}$ and $\alpha$
are two Lorentz-violating parameters that do not contain $\mg$
\cite{Mirshekari+11,Cembranos+18}.
In this paper, we focus on having an effective massive graviton term only,
i.e., we only consider $\omega_\cut = \mg c^2/\hbar$, and numerically studying the resulting modified gravitational wave (GW) energy spectra.
However, we believe the idea of using the GW energy spectra as a
constraint can be applied for a general, Lorentz-violating, modification.

So far, the graviton mass $\mg$ has been constrained
to be $\mg\lsim7.7\times10^{-23}\eV/c^2$ by LIGO
\cite{GW150914,GW170104}, which is already tighter than the constraint on the photon mass $m_\gam\lsim10^{-18}\eV/c^2$ \cite{PDG20}.
However, motivations to develop massive gravity theories remain.
One such motivation is that massive gravity can explain the accelerated
expansion of the universe more ``naturally'' than dark energy, i.e.,
a Yukawa-type gravitational potential of the form $\propto r^{-1}e^{-\mg r c/\hbar}$
\cite{Finn+01,Lee14,Kahniashvili+14}
arises naturally from many massive gravity theories and thus dilutes
the gravitational strength at large distances without the need for
a cosmological constant $\Lam$ \cite{Dvali+02,Babak+02}.
For recent reviews, see ref.~\cite{Maggiore+16} on massive gravity in
the context of $\Lam$-related topics, and ref.~\cite{deRham14} on massive
gravity theories more comprehensively.

Phenomenologically, GWs offer a clean and direct constraint on the graviton mass, albeit not the tightest ones \cite{Desai17}.
Some of the methods using GWs include orbital decay of pulsars \cite{Finn+01},
modified dispersion relation and alternative polarization modes using
pulsar timing arrays \cite{Lee14}, waveforms of extreme-mass-ratio
inspirals \cite{Cardoso+18} and black hole ringdowns \cite{Dong+20}, modified dispersion relation \cite{Barausse+20} and standard sirens
\cite{Belgacem+19} using LISA.
Multi-messenger detection of GW and electromagnetic (EM) waves from
binary neutron star mergers provide another direct constraint on the
propagation speed difference between GW and EM waves.
However, so far the bound using this method is no better than waveform
measurements of GWs alone \cite{Shoemaker+17}; see ref.~\cite{deRham+16}
on various graviton mass bounds.

Although modified GW energy spectra have been studied in the context of
inflation \cite{Fujita+18}, it is not well explored with a turbulent source.
In particular, it is then not clear whether the spectral GW energy is enhanced
or decreased in the presence of a finite graviton mass.
One particular point of uncertainty arises from the fact that the
equivalence between temporal and spatial spectra is now broken.
In turbulence theory and numerical simulations, one computes spatial
spectra, but the measurements in wind tunnels is almost always based on
temporal spectra.
In that case, the approximate equivalence between both spectra
is accomplished by the fact that a chunk of turbulence passes by the
detector at a certain mean speed $\overline{\bm{u}}$, so the fluctuations
as a function of time $t$ can be translated into a spatial dependence
on $\bm{x}$ through $\bm{x}=\bm{x}_0-\overline{\bm{u}} t$, relative to
some reference point $\bm{x}_0$.
Relic GWs, on the other hand, come from all directions, so the equivalence
between spatial and temporal spectra is not that obvious---especially
when there is dispersion.

Meanwhile, significant progress has been made in the numerical solution
of relic GWs from turbulent sources \cite{Pol+18,Pol+19,Kahniashvili+20,Brandenburg+21a,Brandenburg+21}.
A useful tool is the {\sc Pencil Code}, a massively parallel
public domain code developed by the community of users for a broad
range of applications \cite{pencil}.
It comes with a GW solver, where the modification to dispersive
GWs is straightforward.
Therefore, in this work we consider turbulence-sourced GWs in a generic
massive gravity theory by adding a nonzero graviton mass term $\mg$
to the otherwise massless GW equation and explore its effect on the resulting GW spectra.
Specifically, we consider GWs sourced by fully developed turbulence with
an initial Kolmogorov scaling during the electroweak phase transition
(EWPT) \cite{Brandenburg+17} and more realistic hydromagnetic turbulence
that may have been present during the QCD phase transition (QCDPT)
\cite{Brandenburg+21}.
We then compute the resulting energy and strain spectra in spatial and temporal Fourier domains, which are now different due to the dispersion relation being nonlinear.

We begin by briefly recalling the basic phenomenology of
massive GWs (\Sec{sec:massive_gravity_and_its_phenomenology}) and
the relevant parameters for GWs produced during EWPTs and QCDPTs
(\Sec{sec:gravitational_waves_from_the_early_universe}).
We then present the governing equations solved in this
paper (\Sec{sec:hydromagnetic_turbulence_sources}),
and turn then to the discussion of our results
(\Sec{sec:energy_spectra_from_numerical_simulations}).
We conclude in \Sec{sec:discussions_and_conclusions}.
We use the $(-+++)$ metric signature and set $c = 1$, unless specifically
noted otherwise.
We also normalize the critical energy density at the time of GW generation $t_*$ to be unity, i.e., $\RHO_\crit(t_*)= 1$.

\section{Massive gravity and its phenomenology}
\label{sec:massive_gravity_and_its_phenomenology}

For a metric $g_\munu = \eta_\munu + h_\munu$, where $\eta_\munu =
\diag(-+++)$ is the Minkowski background and $|h_\munu|\ll|\eta_\munu|$
is some small perturbation, the action of a generic massive gravity
theory then can be written as
\begin{equation}
S = \int d^4x(\lll_\EH + \lll_{\mg} + \lll_\mat),
\label{eqn:action}
\end{equation}
where $\lll_\EH$ is the usual Einstein-Hilbert Lagrangian in GR,
$\lll_{\mg}$ is a Lagrangian containing the graviton mass term $\mg$, and $\lll_\mat$ is a Lagrangian for matter-energy.
They take the forms
\begin{equation}
\lll_\EH = \sqrt{-g}R,\;\;
\lll_{\mg} = \frac{1}{4}\mg^2\sqrt{-g}\Big(h_\munu h^\munu - \frac{1}{2}h^2\Big),\;\;
\lll_\mat = -8\pi G\sqrt{-g}h^\munu T_\munu,
\label{eqn:lagrangians}
\end{equation}
where $R$ is the Ricci scalar and $T_\munu$ is the stress-energy tensor.

Note that the modified action we are working with here is the Fierz-Pauli (FP) action \cite{FP39}, which is the simplest massive gravity theory. Generalizing the FP action leads to the full dRGT gravity \cite{deRham+10,dRGT10}, for which the massive graviton term reads $\lll_{\mg}\propto\uuu(g,H)$, where $\uuu(g,H)$ is the sum of up to quintic order interaction terms in the perturbation metric $H_\munu$. Conversely, the dRGT gravity reduces to FP gravity at the leading order. Therefore, since our main objective here is to deliver a qualitatively modified GW spectrum instead of the exacts of massive gravity theory, we choose the simpler FP action here.

For a conserved source, the action in equation \eqref{eqn:action} can be
minimized by setting its variation with respect to the metric to be zero,
which reads
\begin{equation}
0 = \frac{\d S}{\d g^\munu} = \int d^4x
\sqrt{-g}\Big(G_\munu + \frac{1}{2}\mg^2\bar h_\munu - 8\pi G T_\munu\Big),
\label{eqn:variation}
\end{equation}
where $G_\munu\eq R_\munu - \frac{1}{2}g_\munu R$ is the Einstein
tensor and $\bar h_\munu\eq h_\munu - \frac{1}{2}\eta_\munu h$ is the
trace-reversed perturbation.
Keeping only the leading order terms in the perturbation $h_\munu$
and applying the harmonic gauge $\p^\mu\bar h_\munu = 0$, the Einstein
tensor becomes
\begin{equation}
G_\munu = \frac{1}{2}\Big(\p_\mu\p_\rho\bar h^\rho_\nu + \p_\nu\p_\rho\bar h^\rho_\mu - \p_\rho\p^\rho\bar h_\munu - \eta_\munu\p_\rho\p_\sig\bar h^{\rho\sig}\Big) 
= -\frac{1}{2}\Box\bar h_\munu,
\label{eqn:einstein_tensor}
\end{equation}
where $\Box\eq\p_\rho\p^\rho = -\p_t^2 + \DDel$ is the d'Alembert operator.

Next, inserting equation \eqref{eqn:einstein_tensor} into equation
\eqref{eqn:variation} gives the linearized equation for massive gravity
\cite{Finn+01},
\begin{equation}
(\Box - \mg^2)\bar h_\munu = -16\pi GT_\munu,
\label{eqn:gw_massive}
\end{equation}
which is the modified GW equation.
Note that we have not yet taken into account the scale factor $a$ characterizing the expansion of the universe -- this will be presented in \Sec{sec:gravitational_waves_from_the_early_universe}.
We drop the overbar of $\bar h_\munu$ from here on.

In momentum space, equation \eqref{eqn:gw_massive} becomes
\begin{equation}
\ddot h_\munu(\kk,t) + (\kk^2 + \mg^2)h_\munu(\kk,t) = 16\pi G T_\munu(\kk,t),
\label{eqn:gw_massive_k}
\end{equation}
where the double dots denote a second derivative in time, i.e.,
$\ddot h_\munu(\kk,t)\equiv\p_t^2 h_\munu(\kk,t)$.
A direct consequence of a generic massive gravity theory is the modified dispersion relation, which now takes the form
\begin{equation}
\omega = \sqrt{k^2 + \omega_\cut^2},
\label{eqn:dispersion_massive}
\end{equation}
where $\omega_\cut = \mg c^2/\hbar$ is an effective mass term for the graviton \cite{Lee14}.

Another important implication of massive gravity
is that the FP action given by equations 
\eqref{eqn:action} and \eqref{eqn:lagrangians} 
propagates five degrees of freedom, 
namely two tensor, two vector, and one scalar mode,
whereas standard GR only contains two tensor modes, $+$ and $\times$.  
For more detailed studies on the polarization states of massive gravity, see ref.~\cite{Tachinami+21}.
In this paper, however, we only study the spectral modifications
of the tensor mode for the following reasons. 
First, detection of any extra polarization modes 
would automatically indicate a modified gravity theory, whereas modifications of the tensor modes 
that are also present in GR 
are more subtle to discern. 
Second, although the construction is different from the FP gravity here, 
there exists a Lorentz-violating minimal theory of massive gravity (MTMG) 
that only carries two tensor modes of GWs \cite{DeFelice+15a,DeFelice+15b}.
These highlight the significance of studying tensor GWs separately.
And third, the nonlinear dispersion in equation
\eqref{eqn:dispersion_massive} holds for all modes, which means the
spectral modifications of tensor modes should have qualitatively valid
features for additional modes, too.
However, perhaps for a future project, it is also important to investigate
the extra polarization modes in the context of low-frequency GWs, since
below around $10^{-7}\Hz$, these modes can have amplitudes similar to
those of the tensor modes in GR \cite{dePaula+04}.

\section{Tensor mode gravitational waves from the early universe}
\label{sec:gravitational_waves_from_the_early_universe}

Equation \eqref{eqn:gw_massive} gives the linearized equation for massive gravity. 
For the early universe, we adopt the normalization such that the
scale factor at the time of GW generation, $t_*$, is set to be unity,
i.e., $a_*\eq a(t_*) = 1$.
The Hubble parameter at that time is $H_\ast$, defined to be
$H_* = \sqrt{8\pi G\RHO_\crit(t_*)/3}$, where $\RHO_\crit(t_*)$
is the critical energy density at time $t_*$.
We denote $\omega_0^H = 2\pi H_*/a_0$ as the Hubble frequency
scaled to the present, $\mg^H = \hbar\omega_0^H$ as the graviton
mass corresponding to the Hubble frequency and $\lambar_0^H = 1/\omega_0^H$
as the reduced Compton wavelength of such a graviton scaled to the present.

Furthermore, we assume an adiabatic expansion of the universe during the
radiation-dominated era.
This means $g_S(T)\,T^3a^3(T)$ stays constant by entropy conservation,
where $g_S(T)$ is the number of adiabatic degrees of freedom at temperature $T$.
Using $T_*^\EW\approx 100\GeV$ and $g_S(T_*^\EW)\approx 100$ at EWPT,
$T_*^\QCD\approx 150\MeV$ and $g_S(T_*^\QCD)\approx 15$ at QCDPT,
combined with $T_0 = 2.7\Kel$ and $g_S(T_0) = 3.9$ at the present, we
obtain $a_0/a_*^\EW\sim1.3\times10^{15}$ and $a_0/a_*^\QCD\sim10^{12}$.

To obtain the Hubble parameter, we need the critical energy density in physical units.
During the radiation era, the critical energy density
of the universe is approximately the radiation energy
density, i.e., $\RHO_\crit(t_*)\simeq\eee_\rad(t_*) =
\pi^2g_S(T_*)k_B^4T_*^4/(30\hbar^3c^3)$, where $\kB$ is the Boltzmann constant.
We note that, in agreement with some earlier work
\cite{Kahniashvili+20,Brandenburg+21a,Brandenburg+21}, we use now the
symbol $\eee$ for mean energy densities, but keep the commonly used symbol
$\RHO$ for the critical density, which includes the rest mass density.

For EWPT and QCDPT, we get $H_*^\EW\approx2.1\times10^{10}\rm s^{-1}$
and $H_*^\QCD\approx1.8\times10^4\rm s^{-1}$, respectively.
Given $H_0\approx3.2\times10^{-18}\rm s^{-1}$ today,
we obtain $H_0/H_*^\EW\approx1.5\times10^{-28}$ and
$H_0/H_*^\QCD\approx1.8\times10^{-22}$.
Using these, we can obtain the energy dilution factor
$(a_*/a_0)^4(H_*/H_0)^2$, which we should multiply the generated
energy by in order to obtain its present day value in the form
$h_0^2\Omega_\GW$ \cite{Pol+18, Pol+19}, which is independent of
the uncertainty in the normalized present day Hubble parameter
$h_0\approx0.7$.
Table \ref{table:scaling_factors} summarizes the aforementioned scaling
factors, together with some useful parameters for EWPT and QCDPT.

We adopt the conformal scaling such that $h_\munu = ah_\munu^\phys$ and
$T_\munu^\TT = a^4 T_{\mu\nu,\phys}^\TT$, and the relation $a(t)\propto t$
during the radiation dominated era, so the GW equation \eqref{eqn:gw_massive}
becomes
\begin{equation}
\Big(\p_t^2 + \mg^2 - \DDel\Big)h_\munu = \frac{16\pi G}{a}T_\munu^\TT,
\end{equation}
where the superscript TT denotes the transverse-traceless (TT) projection.
We solve this equation in Fourier space and work with the stress projected
onto the linear polarization basis, where $+$ and $\times$ denote the
plus and cross polarizations; see \cite{Maggiore19,Caprini+03} for
details and \cite{Pol+18} for the implementation in the {\sc Pencil Code}.

Using normalized conformal time $\bar t = t/t_*$, scaled wave vector
$\bar{\kk} = \kk/H_*$, and scaled normalized stress $\bar{T}^\TT_{+/\times}
= T^\TT_{+/\times}/\eee_\rad^*$, the GW equation in momentum space
(equation \eqref{eqn:gw_massive_k}) becomes
\begin{equation}
\Big(\p_{\bar t}^2 + \mg^2 + \bar{\kk}^2\Big)h_{+/\times}(\kk, t) 
= \frac{6}{\bar t}\,\bar{T}^\TT_{+/\times}(\kk, t),
\label{eqn:gw_massive_k_scaled}
\end{equation}
but we will drop the overbars from now on,\footnote{Note that the overbars
here indicate scaled quantities, which means they are different from
the overbars that appeared, and were subsequently dropped, in equations
\eqref{eqn:variation}--\eqref{eqn:gw_massive}, where 
$\bar{h}_\munu$ denoted the trace-reversed perturbation.}
except for one case
where we explicitly compare with $\xx$ and $t$ in physical space.
Note that this is modified upon equation (13) in ref.~\cite{Pol+18}.

Finally, in terms of observables, we define the characteristic
strain amplitude as $h_\rms = \bra h^2\ket^{1/2}$, where
$h^2 \equiv h_+^2 + h_\times^2 = h_\munu^2/2$, as well as the
scaled GW energy $\eee_\GW = \bra\dot h^2\ket/6$, where
$\dot h^2 = \dot h_+ + \dot h_\times$ and
$\dot h_{+/\times} = \p_t h_{+/\times}$, but see ref.~\cite{Pol+18}
for a small correction term that will here be neglected.

\begin{table}[t]
\centering
\begin{tabular}{|c|c|c|c|c|c|}
\hline
Event & $\kB T$ & $\omega_0^H[\rm Hz]$ & $\mg^H[\eV/c^2]$ & $\lambar_0^H[\rm m]$ & $(a_*/a_0)^4(H_*/H_0)^2$ \\\hline
EWPT  &$100\GeV$& $1.0\times10^{-4}$ & $6.8\times10^{-20}$ & $1.9\times10^{13}$ & $1.6\times10^{-5}$ \\\hline
QCDPT &$150\MeV$& $1.1\times10^{-7}$ & $7.4\times10^{-23}$ & $1.7\times10^{16}$ & $3.1\times10^{-5}$ \\\hline
\end{tabular}
\caption{Scaling factors and useful parameters.}
\label{table:scaling_factors}
\end{table}

\section{Hydromagnetic turbulent sources}
\label{sec:hydromagnetic_turbulence_sources}

The full set of governing equations for the density $\rho$, velocity
field $\uu$, and magnetic field $\BB$ with $\Deldot\BB = 0$ in conformal
time and comoving variables \cite{Brandenburg+96,Brandenburg+17} are
\begin{align}
\pfrac{\ln\rho}{t} = & -\frac{4}{3}(\Deldot\uu + \uu\cdot\Del\ln\rho) + \frac{1}{\rho}[\uu\cdot(\JJ\cdot\BB) + \eta\JJ^2]
\label{eqn:continuity}\\
\begin{split}
\Dfrac{\uu}{t} = & \frac{\uu}{3}(\Deldot\uu + \uu\cdot\Del\ln\rho) - \frac{\uu}{\rho}[\uu\cdot(\JJ\times\BB) + \eta\JJ^2] \\
& - \frac{1}{4}\Del\ln\rho
+ \frac{3}{4\rho}\JJ\times\BB + \frac{2}{\rho}\Deldot(\rho\nu\bm\ssf)
\label{eqn:velocity}
\end{split}\\
\pfrac{\BB}{t} = & \Delcrs(\uu\times\BB - \eta\JJ + \bm\fff),
\label{eqn:mag_evolution}
\end{align}
where $\JJ = \Delcrs\BB$ is the magnetic current,
$D/Dt = \p/\p t + \uu\cdot\Del$ is the advective derivative,
$\nu$ is the kinematic viscosity, $\eta$ is the magnetic diffusivity,
and $\ssf_{ij} = \frac{1}{2}(u_{i,j} + u_{j,i}) - \frac{1}{3}\d_\ij u_{k,k}$
are the components of the traceless strain tensor $\bm\ssf$.

For the EWPT, we start with a turbulence spectrum that has Kolmogorov
scaling in $k$ space, i.e., the initial condition of the evolving turbulence
considered in ref.~\cite{Brandenburg+17}.
We have a magnetic field of the form
\begin{equation}
\BB_i(\kk) = \BB_*[P_\ij(\kk) - i\sig_{\rm M}\eps_\ijl\hat k_l]g_j(\kk)S(k),
\label{eqn:mag_forced}
\end{equation}
where $P_\ij = \d_\ij - \hat{k}_i\hat{k}_j$ is the projection operator,
$\sig_\M$ indicates helicity and is set to 1 in our runs,
$\eps_\ijl$ is the Levi-Civita symbol,
$g_j(\kk)$ is the Fourier transform of a random $\d$-correlated vector field with
Gaussian fluctuations, i.e., $g_i(\xx)g_j(\xx') = \d_\ij\d^3(\xx-\xx')$,
and $S(k)$ determines the spectral shape with
\begin{equation}
S(k) = \frac{\kf^{-3/2}(k/\kf)^{\alpha/2-1}}{[1 + (k/\kf)^{2(\alpha + 5/3)}]^{1/4}},
\label{eqn:mag_shape}
\end{equation}
where $\kf$ is the wave number of the energy-carrying eddies
and $\alpha=4$ for a causal spectrum, such that $S(k)\sim k$ for small $k$ and $S(k)\sim k^{-5/3}$ for large $k$.

For QCDPT, we start with zero magnetic field and, as in
ref.~\cite{Brandenburg+21}, apply instead a forcing function $\bm\fff$ with
\begin{equation}
\bm\fff(\xx,t) = \Re[\nnn\tilde\ff(\kk)\exp(i\kk\cdot\xx+i\varphi)],
\end{equation}
where the wave vector $\kk(t)$ and the phase $\varphi(t)$ change randomly
at each time step.
This forcing function is therefore white noise in time and consists
of plane waves with average wave number $\kf$ such that $|\kk|$
lies in an interval $\kf - \d k/2\leq|\kk|<\kf+\d k/2$ of width
$\d k$.
$\nnn = {\cal F}_0/\d t^{1/2}$ is a normalization factor, where $\d t$
is the time step and ${\cal F}_0$ is varied to achieve a certain
magnetic field strength after a certain time, and $\tilde\fff(\kk) =
(\kk\times\ee)/[\kk^2-(\kk\cdot\ee)^2]^{1/2}$ is a nonhelical forcing
function.
Here $\ee$ is an arbitrary unit vector that is not aligned with $\kk$.
Note that $|\fff|^2=1$.
As in ref.~\cite{Brandenburg+21}, this forcing is only enabled
during the time interval $1\leq t\leq2$.
The kinetic and magnetic energy densities are defined as
$\eee_\K(t) = \bra\rho\uu^2\ket/2$ and $\eee_\M(t) = \bra\BB^2\ket/2$.

\section{Energy spectra from numerical simulations}
\label{sec:energy_spectra_from_numerical_simulations}

We use the {\sc Pencil Code} \cite{pencil} to solve equation~\eqref{eqn:gw_massive_k_scaled}
together with a forced magnetic field given by equations~\eqref{eqn:mag_forced}
and \eqref{eqn:mag_shape} at EWPT, and together with
equations~\eqref{eqn:continuity}--\eqref{eqn:mag_evolution}
for turbulence at QCDPT.
The runs discussed in this paper are summarized in \Tab{table:run_analysis}.
The numerical data for our spectra are publicly available \cite{DATA}.
We recall that in the code, our nondimensional wave numbers and frequencies
correspond, at the present time, to $1/\lambar_0^H$ and $\omega_0^H$,
respectively.
The numerical resolution for the runs is arranged as follows: all EWPT
runs (E1 through F5) have $1152^3$ mesh points each, except for F1, which
has $1024^3$ points; and all QCDPT runs (Q1 through P5) have $512^3$
mesh points. We also set $\nu = \eta = 5\times10^{-5}$ for all runs.
In \Tab{table:run_analysis} we list four groups of runs: two that are applied to the EWPT (E1--E4 with $\eee_{\rm M}=7.8\times10^{-3}$
and F1--F5 with $\eee_{\rm M}=5.6\times10^{-3}$) and two that are
applied to the QCDPT (Q1--Q4 with $\eee_{\rm M}=3.9\times10^{-2}$
and P1--P5 with $\eee_{\rm M}=3.8\times10^{-2}$).
For each group, we vary the value of $\omega_\cut$, which does not
affect the values of $\eee_{\rm M}$, which are therefore not listed
in \Tab{table:run_analysis}.
\begin{table}[t]
\centering
\begin{tabular}{|c|c|c|c|c|c|c|c|}
\hline
Runs & $k_1$ & $\kf$ & $\omega_\cut$ & $\eee_\GW^\sat$ & 
$h_\rms^\sat$ & $h_0^2\Omega_\GW$ & $\hc$ \\
\hline
E1&$100$&$600$&  0&$1.15\times10^{-10}$&$9.36\times10^{-8}$&$1.89\times10^{-15}$&$7.46\times10^{-23}$\\
E2&$100$&$600$& 10&$1.13\times10^{-10}$&$1.15\times10^{-7}$&$1.85\times10^{-15}$&$9.19\times10^{-23}$\\
E3&$100$&$600$& 50&$1.12\times10^{-10}$&$9.95\times10^{-8}$&$1.85\times10^{-15}$&$7.93\times10^{-23}$\\
E4&$100$&$600$&200&$1.03\times10^{-10}$&$6.44\times10^{-8}$&$1.69\times10^{-15}$&$5.13\times10^{-23}$\\
\hline
F1&$ 1 $&$100$& 0 &$3.77\times10^{-9}$&$8.93\times10^{-6}$&$6.20\times10^{-14}$&$7.12\times10^{-21}$\\
F2&$ 1 $&$100$&0.3&$3.84\times10^{-9}$&$8.34\times10^{-6}$&$6.31\times10^{-14}$&$6.65\times10^{-21}$\\
F3&$ 1 $&$100$& 1 &$3.83\times10^{-9}$&$7.06\times10^{-6}$&$6.29\times10^{-14}$&$5.63\times10^{-21}$\\
F4&$ 1 $&$100$& 3 &$3.74\times10^{-9}$&$3.91\times10^{-6}$&$6.14\times10^{-14}$&$3.12\times10^{-21}$\\
F5&$ 1 $&$100$&10 &$3.68\times10^{-9}$&$2.01\times10^{-6}$&$6.06\times10^{-14}$&$1.60\times10^{-21}$\\
\hline
Q1&0.3&2& 0 &$4.86\times10^{-4}$&$4.16\times10^{-2}$&$1.50\times10^{-8}$&$4.17\times10^{-14}$\\
Q2&0.3&2&0.3&$4.70\times10^{-4}$&$3.44\times10^{-2}$&$1.45\times10^{-8}$&$3.44\times10^{-14}$\\
Q3&0.3&2& 1 &$3.91\times10^{-4}$&$2.05\times10^{-2}$&$1.21\times10^{-8}$&$2.06\times10^{-14}$\\
Q4&0.3&2& 3 &$1.99\times10^{-4}$&$8.48\times10^{-3}$&$6.15\times10^{-9}$&$8.49\times10^{-15}$\\
Q5&0.3&2&10 &$3.41\times10^{-5}$&$1.37\times10^{-3}$&$1.05\times10^{-9}$&$1.37\times10^{-15}$\\
\hline
P1&1&6& 0 &$5.05\times10^{-5}$&$4.62\times10^{-3}$&$1.56\times10^{-9}$&$4.63\times10^{-15}$\\
P2&1&6&0.3&$5.02\times10^{-5}$&$4.48\times10^{-3}$&$1.55\times10^{-9}$&$4.48\times10^{-15}$\\
P3&1&6& 1 &$4.81\times10^{-5}$&$3.54\times10^{-3}$&$1.49\times10^{-9}$&$3.55\times10^{-15}$\\
P4&1&6& 3 &$3.97\times10^{-5}$&$2.07\times10^{-3}$&$1.23\times10^{-9}$&$2.08\times10^{-15}$\\
P5&1&6&10 &$1.85\times10^{-5}$&$7.71\times10^{-4}$&$5.72\times10^{-10}$&$7.71\times10^{-16}$\\
\hline
\end{tabular}
\caption{Summary of runs shown in the paper. $k_1$ is the smallest wave number
in the simulation domain, and $\kf$ characterizes the peak magnetic energy.
$h_0^2\Omega_\GW$ and $\hc$ are the values of $\eee_\GW^\sat$ and $h_\rms^\sat$
scaled to the present day, respectively.}
\label{table:run_analysis}
\end{table}

\subsection{Spatial Fourier spectra}

Following ref.~\cite{Pol+18}, we define spatial Fourier spectra as
integrals over concentric shells in wave number space
(indicated now by a single tilde), i.e.,
\begin{equation}
\Sp_k(h)=\int_{4\pi}\left[\left|\tilde{h}_+(\kk,t)\right|^2
+\left|\tilde{h}_\times(\kk,t)\right|^2\right]k^2\,\dd\Omega_k,
\label{eqn:Temporal_Spectrum}
\end{equation}
where $\dd\Omega_k$ is the differential over the solid angle in $\kk$ space.
An analogous definition applies also to $\dot{h}$, which is used for
calculating $E_{\rm GW}(k,t)=\Sp_k(\dot{h})/6$, but see equation~(B.36)
of ref.~\cite{Pol+18} for lower order correction terms that are
here neglected.

\begin{figure}[H]
\begin{subfigure}{\textwidth}
\centering
\includegraphics[width=0.95\linewidth]{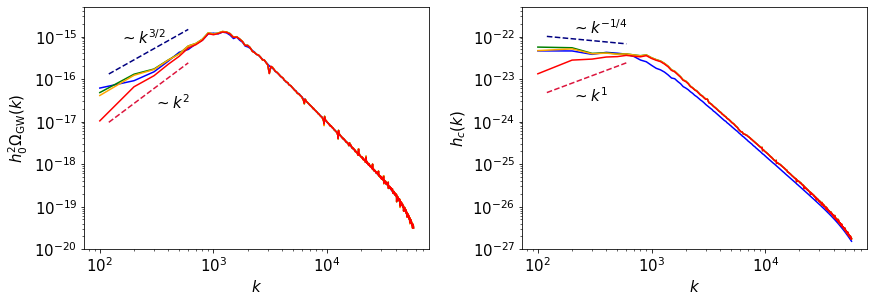}
\caption{Runs E1 to E4: $\omega_\cut = 0$ (blue), 10 (green), 50 (orange), and 200 (red).}
\label{fig:pspecm_hel_delk_kf600} 
\end{subfigure}
\\
\begin{subfigure}{\textwidth}
\centering
\includegraphics[width=0.95\linewidth]{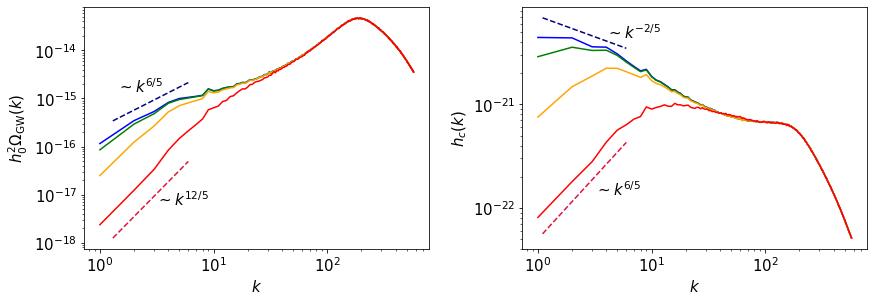}
\caption{Runs F2 to F5: $\omega_\cut = 0.3$ (blue), 1 (green), 3 (orange), and 10 (red).}
\label{fig:pspecm_hel_delk_kf100}
\end{subfigure}
\\
\begin{subfigure}{\textwidth}
\centering
\includegraphics[width=0.95\linewidth]{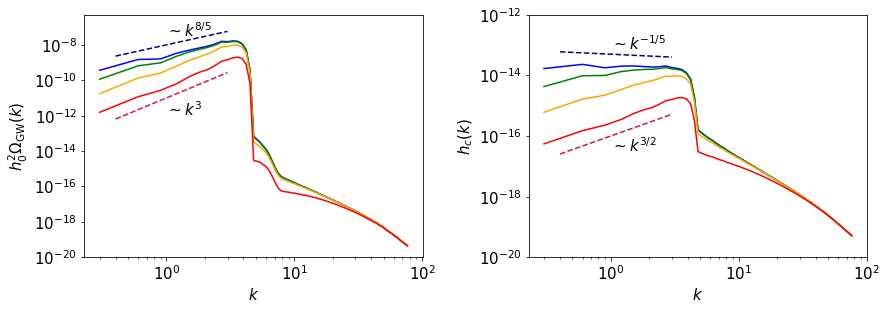}
\caption{Runs Q2 to Q5: $\omega_\cut = 0.3$ (blue), 1 (green), 3 (orange), and 10 (red).}
\label{fig:pspecm_hel_delk} 
\end{subfigure}
\\
\begin{subfigure}{\textwidth}
\centering
\includegraphics[width=0.95\linewidth]{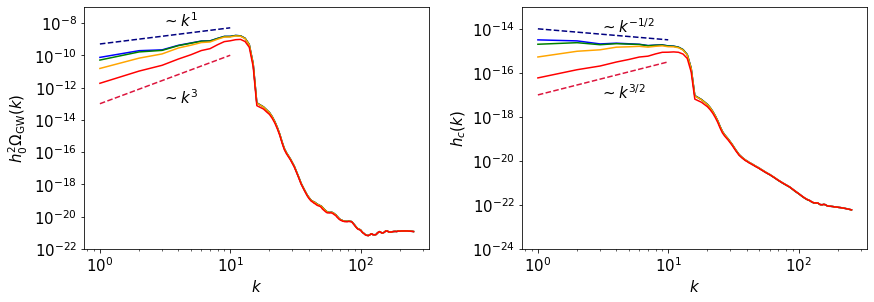}
\caption{Runs P2 to P5: $\omega_\cut = 0.3$ (blue), 1 (green), 3 (orange), and 10 (red).}
\label{fig:pspecm_hel_delk_kf6}
\end{subfigure}
\caption{Spectra of GW energy density $h_0^2\Omega_\GW(k)$ and strain $\hc(k)$ from EWPT and QCDPT scaled to the present time.}
\label{fig:pspecm_hel_delk_qcd}
\end{figure}

For EWPT, we first choose the same parameters as in ref.~\cite{Pol+19},
where the smallest wave number in the simulation
domain is $k_1 = 100$ and the peak wave number is $\kf = 600$.
The resulting GW energy density today, $h_0^2\Omega_\GW(k)$, and the
strain today, $\hc(k)$, as functions of wave number $k$ are shown in
figure~\ref{fig:pspecm_hel_delk_kf600}.
We see that for all the effective mass terms $\omega_\cut\in\{10,50,200\}$
corresponding to $\mg\in\{6.8\times10^{-19},3.4\times10^{-18},1.4\times10^{-17}\}\eV/c^2$,
the spectral modifications are not significant, even though $\omega_\cut = 200$
already corresponds to an unrealistically large graviton mass
$\mg = 1.4\times10^{-17}\eV/c^2$.

Next, we explore the same EWPT era but with $k_1 = 1$ and $\kf = 100$, 
and the effective mass term $\omega_\cut\in\{0.3,1,3,10\}$, 
corresponding to a range of graviton masses
$\mg\in\{2.0\times10^{-20},6.8\times10^{-20},2.0\times10^{-19},6.8\times10^{-19}\}\eV/c^2$. 
The resulting spectra are shown in
figure~\ref{fig:pspecm_hel_delk_kf100}.
Now the spectral differences are more significant than before,
especially towards the lower wave number at around $k_1$.
For the spectral energy density $h_0^2\Omega_\GW(k)$, 
the left panel of figure~\ref{fig:pspecm_hel_delk_kf100} shows
spectral shape changes by about a factor of $k^{1.2}$, from $k^{6/5}$
to $k^{12/5}$.
For the strain $\hc(k)$, 
the right-hand panel of figure~\ref{fig:pspecm_hel_delk_kf100} shows slope 
changes correspondingly by about a factor $k^{1.6}$, from $k^{-2/5}$ to $k^{6/5}$
as we increase the graviton mass.
However, the constraint on the graviton mass provided by GWs from EWPT is
not significant overall, as can be seen from \Tab{table:scaling_factors},
compared to the existing constraint of $\mg\lsim7.7\times10^{-23}\eV/c^2$
\cite{GW150914,GW170104} that we quoted in the introduction.

Since the QCDPT era could provide a constraint on the graviton mass tighter than EWPT by about three orders of magnitude (\Tab{table:scaling_factors}),
we would like to explore the spectral behaviors of GWs from the QCDPT era.
For this we adopt the simulation setup of previous work \cite{Brandenburg+21}, since the only change
here is adding an effective graviton mass term $ \omega_\cut$.
We have two series of runs, the first with $k_1 = 0.3$
and $\kf = 2$ (runs Q2 to Q5), and the second with $k_1 = 1$ and $\kf = 6$ (runs P2 to P5).
For both series, we vary $\omega_\cut\in\{0.3, 1, 3, 10\}$,
corresponding to a range of graviton masses
$\mg\in\{2.2\times10^{-23},7.4\times10^{-23},2.2\times10^{-22},7.4\times10^{-22}\}\eV/c^2$.
The resulting energy and strain spectra can be seen in figures \ref{fig:pspecm_hel_delk} and
\ref{fig:pspecm_hel_delk_kf6}, which show remarkably consistent
modifications for the two series. We see that for both cases, as the
graviton mass increases, the spectral shapes at lower wave numbers,
around the corresponding value of $k_1$,
become steeper. 
In particular, figure \ref{fig:pspecm_hel_delk} shows that for
$\kf = 2$, the energy density $h_0^2\Omega_\GW$ goes from $k^{8/5}$
to $k^{3}$, and the strain $\hc$ goes from $k^{-1/5}$ to $k^{3/2}$;
and figure \ref{fig:pspecm_hel_delk_kf6} shows that for $\kf = 6$, 
the energy density changes from $k^1$ to $k^3$, 
and the strain from $k^{-1/2}$ to $k^{3/2}$, respectively.
Therefore, the spectral modifications are only weakly dependent on the
driving wave number.

Note that in figures \ref{fig:pspecm_hel_delk_kf600}--d,
we observe the expected slope correspondence between GW energy
density and strain, i.e., in the small graviton mass limit, we have
$h_0^2\Omega_\GW(k)\propto k^2 h_c^2(k)$, and in the large graviton mass
limit, the relation changes to $h_0^2\Omega_\GW(k)\propto h_c^2(k)$.

To understand the systematic change from $h_0^2\Omega_\GW(k)\propto k^3$
to $\hc(k)\propto k^{3/2}$ in the QCD runs, it is important to recall that for a causal
spectrum of the magnetic field with $\Sp(\BB)\propto k^4$ \cite{DC03},
the stress only has a spectrum proportion to $k^2$ \cite{BB20}.
This is because a blue spectrum convolved with itself can only become
white noise.
It implies that $\Sp([k^2 + \omega_\cut^2]h)\propto k^2$, and therefore
\begin{equation}
\hc=\sqrt{k\,\Sp(h)}\propto k^{3/2}/(k^2 + \omega_\cut^2) \propto k^{3/2}
\end{equation}
for $k\ll\omega_\cut$.
The spectrum of $\dot{h}$, on the other hand, is given by
$\omega^2\Sp(h)\propto k^2/(k^2 + \omega_\cut^2)$, and therefore
\begin{equation}
h_0^2\Omega_\GW(k)=k\,\Sp({\dot{h})}\propto k^3/(k^2+\omega_\cut^2)\propto k^3
\end{equation}
for $k\ll\omega_\cut$.
These considerations clearly demonstrate the obtained trend 
in figures \ref{fig:pspecm_hel_delk} and \ref{fig:pspecm_hel_delk_kf6} 
from the
$h_0^2\Omega_\GW(k)\propto k^3$ scaling to a $\hc(k)\propto k^{3/2}$
scaling.
We also see that, relative to the case $\omega_\cut=0$, both spectra
decrease in amplitude by a factor $1/\omega_\cut^2$.
This is well borne out by the simulations, where we see a drop by
$1/100$ in both panels of figure~\ref{fig:pspecm_hel_delk_kf100} for
$\omega_\cut=10$.

For all the graviton masses considered, the patterns at higher wave
numbers are approximately the same, including a sharp drop by many
orders of magnitude found previously in the massless case discussed in
ref.~\cite{Brandenburg+21}.
We return to this feature further below.

In order to make closer connections to potential detections, we would
like to inspect the temporal Fourier spectra next.
In particular, we will show that both EWPT and QCDPT could induce
observable features in the $\sim10\nHz$ range, accessible by NANOGrav.

\subsection{Temporal Fourier spectra}
\label{sec:Temporal_Fourier_spectra}

\begin{figure}[H]\begin{center}
\includegraphics[width=\textwidth]{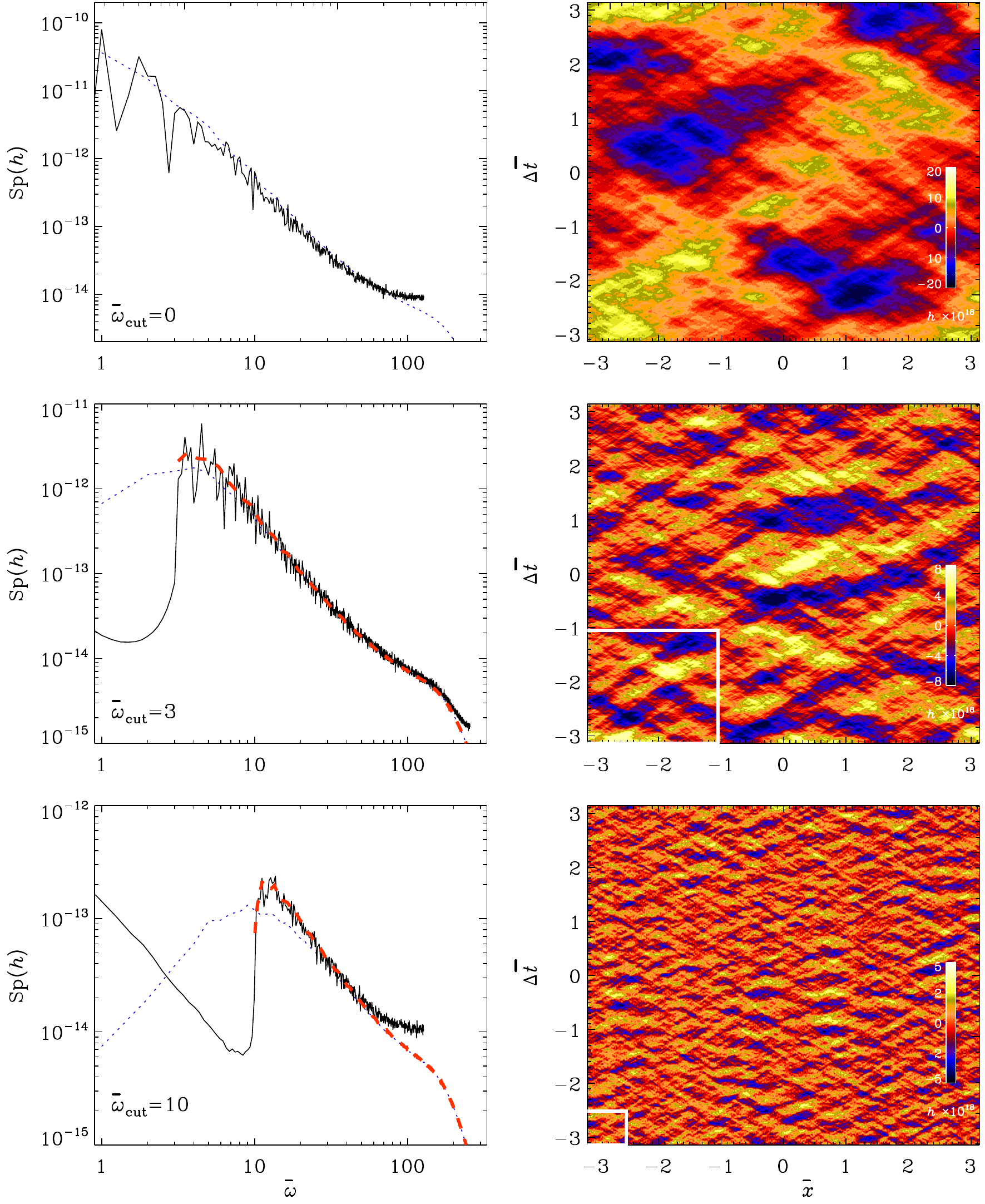}
\end{center}\caption[]{
Left: temporal strain spectra (solid lines) compared with spatial
strain spectra with (red dashed lines) and without (blue dotted lines)
the transformation to $\omega$ space applied, for $\omega_{\rm cut}=0$,
3, and 10.
Right: space--time diagrams of the strain
for runs~F1, F4, and F5 with $\omega_{\rm cut}=0$, 3, and 10.
The small white boxes on the lower left corner of the last two strain
diagrams with a size of $2\pi/\omega_\cut$ are given for orientation.
The color bars give the present-day physical strain multiplied by $10^{18}$.
}\label{fig:pcomp_last}\end{figure}

As alluded to in the introduction, spatial and temporal spectra can be
different from each other if there is nonlinear dispersion.
To the best of our knowledge, it is the first time that GW spectra
have been obtained from turbulence
simulations in terms of temporal frequency.
To compute temporal spectra, we first Fourier transform $\tilde{h}_+(\kk,t)$
and $\tilde{h}_\times(\kk,t)$ back into real space to obtain time series
\begin{equation}
h_{+/\times}(\xx,t)=\int\tilde{h}_{+/\times}(\kk,t)\,
e^{\ii\kk\cdot\xx}\,\dd^3\kk/(2\pi)^3.
\end{equation}
We then compute their Fourier transforms to $\omega$ space as
$\tilde{\tilde{h}}_{+/\times}(\xx,\omega)=\int h_{+/\times}(\xx,t)\,
e^{\ii\omega t}\,\dd t$ at several points $\xx_i$.
We finally compute the mean spectrum as
\begin{equation}
\Sp_\omega(h)=N^{-1}\sum_{i=1}^N\left[\left|\tilde{\tilde{h}}_+(\xx_i,\omega)\right|^2
+\left|\tilde{\tilde{h}}_\times(\xx_i,\omega)\right|^2\right],
\label{eqn:Temporal_Spectrum}
\end{equation}
which is an average over $N$ spatial points.
In practice, we take $N=1024$, which is the number of mesh points of
the simulations in the $x$ direction.
The results for $\Sp_\omega(h)$ are shown in \Fig{fig:pcomp_last},
where we also compare with $\Sp_k(h)$ using both $\omega=k$
(ignoring dispersion, which is only valid for $\omega_\cut=0$)
and $\omega=\sqrt{k^2+\omega_\cut^2}$ (valid also in the presence
of dispersion, with $\omega_\cut\neq0$).
Since
\begin{equation}
\int\Sp_\omega(h)\,\dd\omega=\int\Sp_k(h)\,\dd k=h_{\rm rms}^2,
\label{eqn:sp_omega_sp_k}
\end{equation}
we present in the following $(\omega/k)\,\Sp_{k(\omega)}(h)$,
where the factor $\dd k/\dd\omega=\omega/k$ (for $\omega>\omega_\cut$)
has been applied to take the effect of dispersion into account.

In \Fig{fig:pcomp_last}, we also show contour plots of $h(\xx,t)$
through an arbitrarily selected section $\xx=(x,0,0)$ for the same three
cases with $\omega_\cut=0$, 3, and 10.
They would give us a direct impression of how the GW field would vary
in the proximity of the Sun in space and time.

The temporal spectra show a sharp drop near $\omega=\omega_\cut$,
but $\Sp_\omega(h)$ only drops by about two orders of magnitude
and does not vanish completely.
Moreover, towards smaller values of $\omega<\omega_\cut$,
$\Sp_\omega(h)$ begins to rise again.
This is probably just a consequence of having a finite length
of the time series, which makes the statistical error large at
small frequencies.

The drop in spectral power below $\omega_\cut$ also has marked
consequences for the appearance of $h(x,t)$ in real space
(right-hand panels of \Fig{fig:pcomp_last}, which show
smaller scale patches in space and time whose size is comparable
to $2\pi/\omega_\cut$ in space, but only about half
as long as $2\pi/\omega_\cut$ in time.
For $\omega_\cut=0$, on the other hand, we see extended patches in
$h(x,t)$ that are all comparable in size to the scale of the domain.
We note at this point that in all three plots of $h(x,t)$, we have
selected a time interval of the length $2\pi$, i.e., the same
interval as in $x$.
We select the time interval to be near the end of the run, but other
time intervals look similar.

On the outer axes of \Fig{fig:pcomp_last},
we have indicated the values scaled to a physical time corresponding to the QCDPT,
using $t=(a_0/H_\ast)\,\bar{t}$ and $x=(c a_0/H_\ast)\,\bar{x}$ with
$a_0=10^{12}$, $H_\ast=1.8\times10^4\s^{-1}$, $c=3\times10^8\m\s^{-1}$
being the physical dimensions used.
We see that the patches can become as short as a fraction of a year.

\begin{figure}[H]\begin{center}
\includegraphics[width=\textwidth]{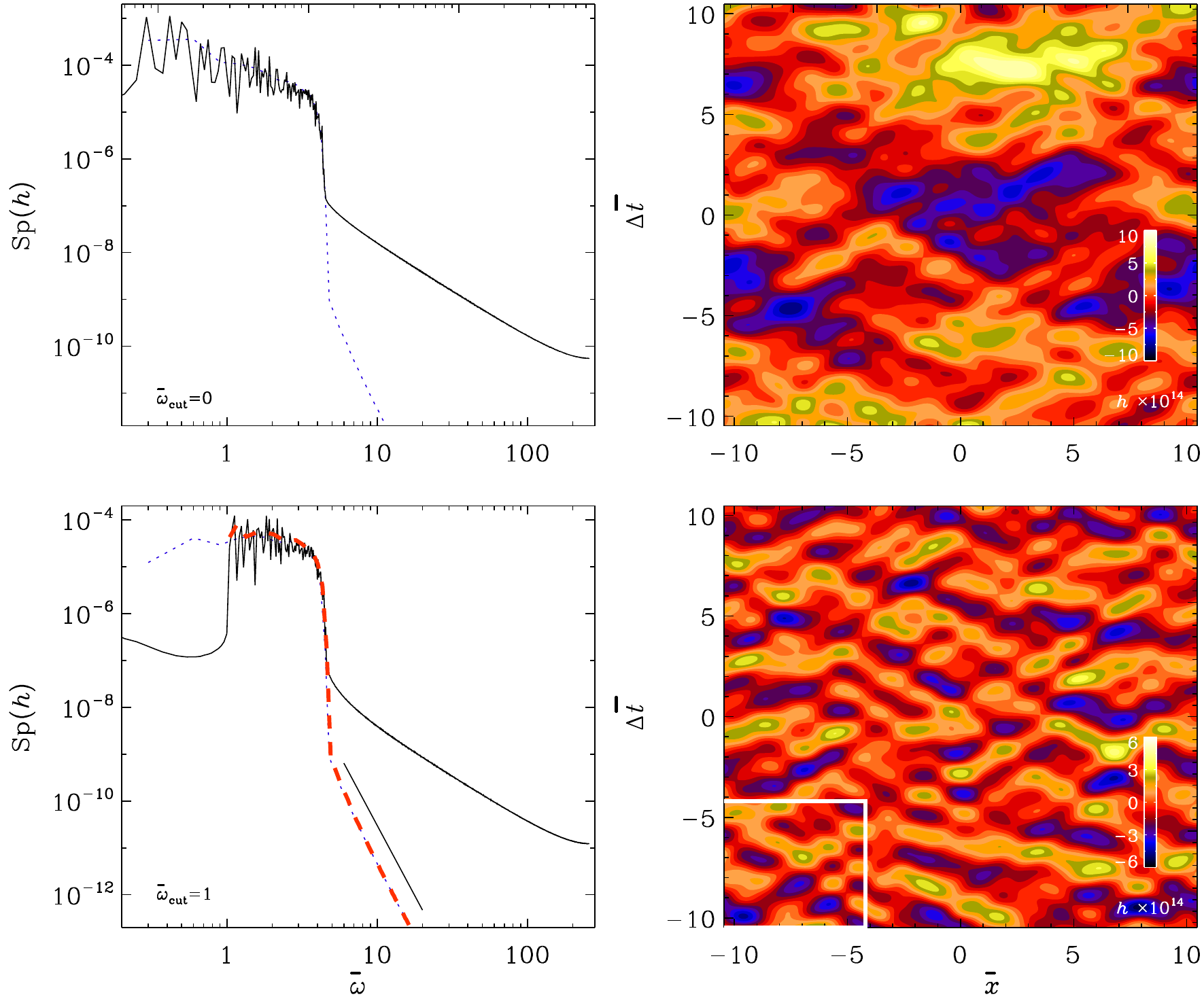}
\end{center}\caption[]{
Similar to \Fig{fig:pcomp_last}, but for runs~Q1 and Q3
with $\omega_{\rm cut}=0$ and 1, respectively.
The color bars give the physical strain multiplied by $10^{14}$.
}\label{fig:pcomp_last_QCD}\end{figure}

In \Fig{fig:pcomp_last_QCD}, we show $\Sp_\omega(h)$ and contour
plots of $h(\xx,t)$ for runs~Q1 and Q3 with $\omega_{\rm cut}=0$ and 1.
One of the major features in the runs shown in \Fig{fig:pcomp_last} is
the presence of a sharp drop of spectral power for wave numbers
above $2\kf\approx6$.
This translates to a similar drop also in the $\omega$ spectra,
but now, when $\omega_{\rm cut}=1$, GWs exist only over a
narrow range of frequencies.
This is probably also the reason why the space--time diagrams
of the strain show a somewhat more regular pattern for
$\omega_{\rm cut}=1$.

Finally, we note that in figure~\ref{fig:pcomp_last_QCD} between
$\omega\sim10$ and $\omega\sim100$, the blue dotted lines sharply drop
by several orders of magnitude, exposing a divergence between the black
solid lines.
This disagreement between the $\omega$ and $\kk$ spectra is not related
to the graviton mass, but it is primarily related to the sharp drop in
$\Sp(h)$ above $\bar{\omega}=\kf=3$.
We have seen such a drop in runs with low $\kf$ for the QCD runs
(figures~\ref{fig:pspecm_hel_delk}
and \ref{fig:pspecm_hel_delk_kf6}).
One might think that
it could be related to the Reynolds number not being large enough.
However, we have been able to increase the Reynolds number by a factor
of ten and have still not seen any significant change in this drop.
On the other hand, temporal spectra are not as accurate as spatial
ones, which take the entire data cube into account.
They are therefore more easily affected by statistical noise.
Given that the drop extends over several orders of magnitude,
the differences between the spatial and temporal spectra
occur where the spatial spectrum is already very weak and
below the noise level of the temporal spectrum.
Note that this feature of a sharp drop is also present
in a physically more realistic turbulence source
driven by inflationary magnetogenesis \cite{Brandenburg+21b}.
The weak temporal signal beyond this drop
is therefore most likely no longer reliable.

\section{Discussions and conclusions}
\label{sec:discussions_and_conclusions}

We have seen that, depending on different values of graviton masses
considered, GW energy and strain can exhibit significant modifications,
mainly in terms of spectral shapes.
The spectra of relic GWs from EWPT and QCDPT have the potential
to constrain the graviton mass and, in turn, massive gravity
theories.
For GWs from fully developed hydromagnetic turbulence at
EWPT, the spectral changes are not so pronounced when
$\kf = 600$ and $k_1 = 100$, but become more
significant when $\kf = 100$ and $k_ 1 = 1$, allowing us to see
changes at low wave numbers near $k_1$.
Specifically, the changes here are that the energy 
and strain spectra both become steeper by slightly more than a $k^1$ factor
as we increase the graviton mass by about 30 fold.
For QCDPT, we see that GWs induced by forced hydromagnetic turbulence
also exhibit steepening spectral slopes around $k_1$ by up to
a $k^2$ factor as the graviton mass increases by about 30 fold.
Quite remarkably, the spectral slopes are only weakly affected by
different driving wave numbers, meaning that, even if we are unsure
about the exact number of turbulent eddies at the time of QCDPT, we
could still use the GW spectra to constrain the graviton mass.
For both EWPT and QCDPT, the relic GW energy in the low wave number tails
is expected to be lower in the presence of a nonzero graviton mass,
although the values at high wave numbers are roughly unchanged.

The frequency spectra can directly be obtained from wave number
spectra through a simple transformation, which takes
$\omega_{\rm cut}\neq0$ into account.
Our work has demonstrated that the agreement between both types
of spectra is rather good, but frequency spectra obtained from
a single detector can be rather noisy.
As the name $\omega_{\rm cut}$ suggests, there is no signal below the cutoff frequency.
Determining this cutoff frequency can be a sensitive means for
constraining the value of the graviton mass.
In wave number space, by contrast, the graviton mass only affects the
spectral slope in the proximity of the associated cutoff wave number.
Even in real space, a finite graviton mass manifests itself through a
striking absence of waves with periods above the cutoff value.

\paragraph{Data availability.} The source code used for the simulations of this study,
the {\sc Pencil Code}, is freely available \cite{pencil};
see also ref.~\cite{DATA} for the numerical data of the spectra.

\acknowledgments
We thank the anonymous referee for useful remarks and suggestions.
Support through the grant 2019-04234 from the Swedish Research Council
(Vetenskapsr{\aa}det) is gratefully acknowledged.
We acknowledge the allocation of computing resources provided by the Swedish National Allocations Committee at the Center for Parallel Computers at the Royal Institute of Technology in Stockholm.


\end{document}